\documentclass[journal]{IEEEtran}
\usepackage{lineno}
\usepackage{cite}
\usepackage{hyperref}
\usepackage{amsmath,amssymb,amsfonts}
\usepackage{amsmath}
\usepackage{amsthm}

\usepackage{graphicx}
\usepackage{textcomp}
\usepackage{xcolor}
\usepackage{graphicx}
\usepackage{float}
\usepackage{subfigure}
\usepackage{amsmath}
\usepackage{amsfonts,amssymb}
\usepackage{mathrsfs}
\usepackage{mathtools}
\usepackage{multirow}
\usepackage{algpseudocode}
\usepackage{hyperref}
\makeatletter
\newif\if@restonecol
\makeatother

\usepackage[linesnumbered,ruled,vlined]{algorithm2e}
\usepackage{algpseudocode}

\usepackage{setspace}
\usepackage{footmisc}
\usepackage[justification=centering]{caption}

\usepackage{mathtools}
\usepackage{dsfont}
\usepackage{bbm}
\usepackage{stfloats}
\usepackage{setspace}
\usepackage{titlesec}
\titlespacing{\subsection} {0pt}{0pt}{0pt}

\newtheorem{corollary}{Corollary}

\theoremstyle{definition}
\newtheorem{theorem}{Theorem}

\newtheorem{lemma}{Lemma}

\makeatletter
\newcommand{\biggg}{\bBigg@{3}}
\newcommand{\Biggg}{\bBigg@{3.5}}
\makeatother
\usepackage{bm}
\makeatother
\begin{document}

\title{Sum-Rate Optimization for RIS-Aided Multiuser Communications with Movable Antenna \\}

\author{Yunan~Sun, Hao~Xu, Chongjun~Ouyang, and Hongwen~Yang
\thanks{The authors are with the School of Information and Communication Engineering, Beijing University of Posts and Telecommunications, Beijing, 100876, China (e-mail: \{sunyunan, Xu\_Hao, DragonAim, yanghong\}@bupt.edu.cn).}
\thanks{(Corresponding author: Hongwen Yang)}
}

\maketitle

\begin{abstract}
Reconfigurable intelligent surface (RIS) is known as a promising technology to improve the performance of wireless communication networks, which has been extensively studied. Movable antenna (MA) is a novel technology that fully exploits the antenna position for enhancing the channel capacity. In this paper, we propose a new RIS-aided multiuser communication system with MAs. The sum-rate is maximized by jointly optimizing the beamforming, the reflection coefficient (RC) values of RIS and the positions of MAs. A fractional programming-based iterative algorithm is proposed to solve the formulated non-convex problem, considering three assumptions for the RIS. Numerical results are presented to verify the effectiveness of the proposed algorithm and the superiority of the proposed MA-based system in terms of sum-rate.

\end{abstract}

\vspace{-10pt}
\section{INTRODUCTION}
With the development of metasurfaces, reconfigurable intelligent surface (RIS) has attracted a great deal of attentions as a promising technology to realize the smart radio environments (SRE)\cite{Liu2021}. An RIS is a two-dimensional surface consisting of many low-cost passive elements and a controller. By adjusting the propagation of the incident signals smartly, RIS is able to improve the performance of wireless communication networks with increasing number of degrees of freedom (DoFs) into the channel, especially when the line-of-sight (LoS) paths between the base station (BS) and user terminals (UTs) are blocked\cite{Guo2019}. Thus, RIS proves to be the most promising technology to enhance the efficiency of data transmission. A lot of research has been done to introduce RIS as an aid to conventional multiuser communication system\cite{Xiu2021}-\cite{Pan2020}.

On the other hand, the concept of movable antennas (MAs) was recently conceived to harness additional spatial DoFs, which can overcome some constraints of conventional fixed-position antennas (FPAs)\cite{Zhu2022}. The positions of MAs can be adjusted by controllers in real time, such as stepper motors or servos, since the MAs are connected to radio frequency (RF) chains via flexible cables. Compared with conventional FPA-based wireless communication systems, the MA-based systems can reasonably change the positions of transmit antennas depending on the channel state information (CSI) such that the channel between MAs and UTs is reshaped for achieving higher capacity\cite{Pi2023}. Moreover, the MAs can be flexibly moved in a three-dimensional (3D) region to fully exploit the channel variation therein\cite{Zhu2023}.Thus the MA-based system can exploit the full spatial diversity in a given region with much fewer antennas than the conventional antenna selection (AS) technique.Due to the superiority of MA, the concept of MA has attracted increasing attention. The recieve signal-to-noise ratio (SNR) of a MA-based system was analyzed in \cite{Zhu2022}. In \cite{Ma2023}, the authors initially characterized the capacity of a point-to-point MIMO channel with MAs, where both the BS and the UTs are equipped with MAs. An fractional programming (FP)-based algorithm was first introduced to an MA-based downlink transmission framework to tackle the joint optimization problem of transmit beamforming and MA positions for maximizing the sum-rate in \cite{Cheng2023}.

As no existing works focus on the MA-based multiuser wireless communication system with the aid of RIS, we investigate an RIS-aided MA enabled downlink multiuser multiple-input single output (MU-MISO) system in this paper, where only the BS is equipped with MAs and the LoS paths between the BS and UTs are blocked. Under this scenario, we focus on the joint optimization problem of beamforming, reflection coefficient (RC) values of RIS and the positions of MAs aiming to maximize the sum-rate of the presented system.

The main contributions of this work is summarized as follows:
\begin{itemize}
  \item This paper is one of the early attempts to study the sum-rate optimization for the RIS-aided MA enabled downlink MU-MISO system. We propose this framework that harness both the RIS and MAs to improve sum-rate.
  \item We propose an iterative FP-based algorithm to solve the joint optimization problem of beamforming, RC values of RIS and the positions of MAs.
  \item Numerical results have verified the effectiveness of the proposed algorithm and the superiority of the proposed MA-based system in terms of sum-rate.
\end{itemize}

\section{SYSTEM MODEL}
\subsection{System Description}
\begin{figure}[!t]
\centering
\setlength{\abovecaptionskip}{0pt}
\includegraphics[height=0.25\textwidth]{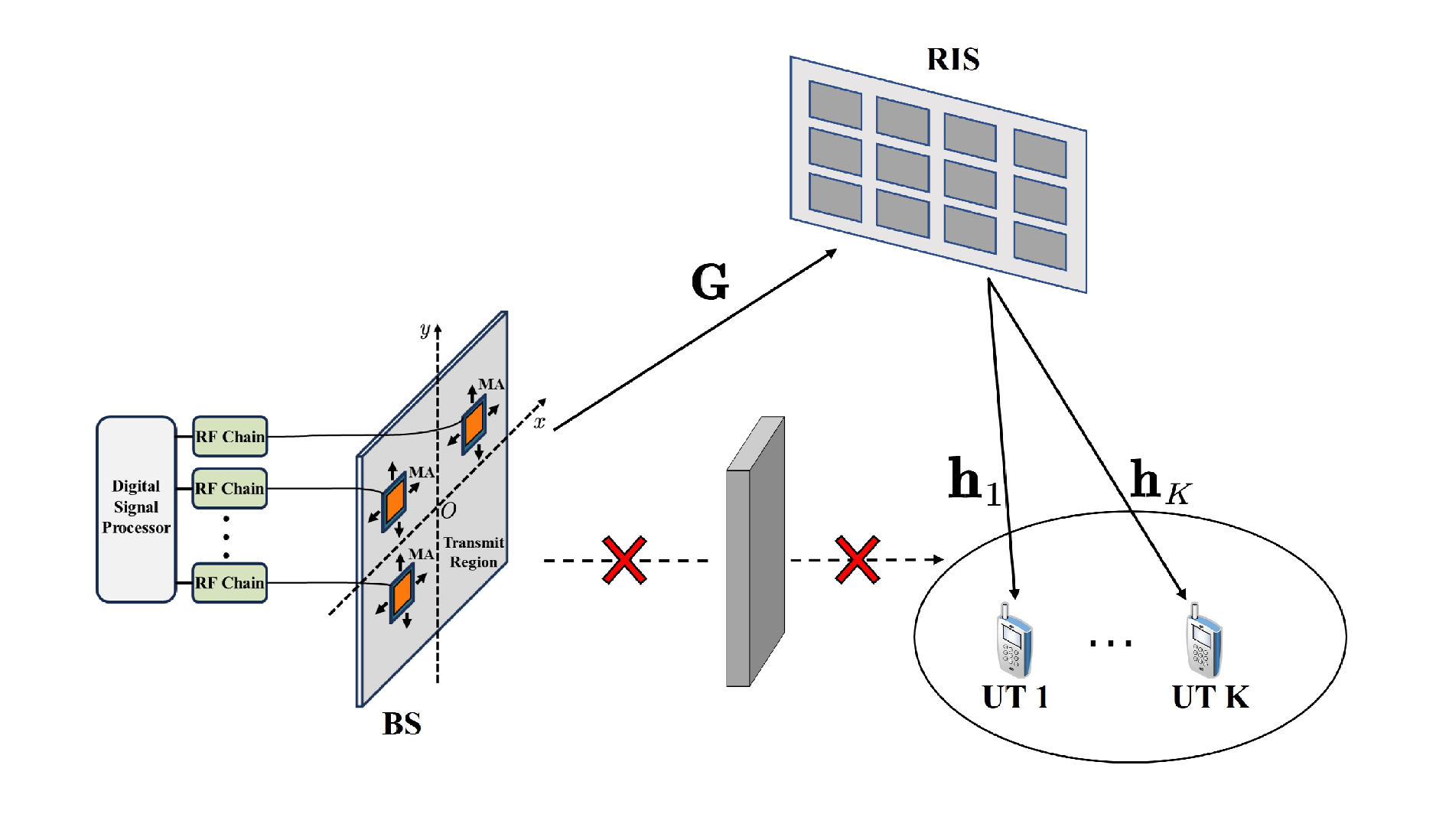}
\captionsetup{font={stretch=1.5}}
\caption{The RIS-aided communication system with MA-BS.}
\label{RIS_MUMISO}
\vspace{-20pt}
\end{figure}

We investigate an RIS-aided multiuser MISO downlink communication system with movable-antenna BS as shown in {\figurename} \ref{RIS_MUMISO}, where the BS equipped with $N$ transmit antennas sends signals to $K$ single antenna UTs with an RIS which has $M$ reflection elements. The UTs are denoted by $\mathcal{K}\triangleq\{1,...,K\}$. Each UT has a receive FPA while the BS has N transmit MAs. All the MAs can move to change their positions in real time since they are connected to RF chains via flexible cables. The position of $n$-th MA is represented by Cartesian coordinates $\textbf t_n=[x_n,y_n]^T \in \mathcal{C}$ for $n \in \mathcal{N}\triangleq\{1,...,N\}$, where $\mathcal{C}$ is a square region with size $A\times A$.

For simplicity, we assume that all the channels experience quasi-static flat-fading. In addition, we assume that the CSI of all channels involved is perfectly known by the BS and the IRS. The RIS-aided link can be modeled as a concatenation of three components, i.e., the BS-RIS link, RIS phase-shift matrix and RIS-UT link. Thus the channel vector $\textbf H_k\in \mathbbmss{C}^{N\times 1}$ from the BS to $k$-th UT follows the structure as:
{\setlength\abovedisplayskip{2pt}
\setlength\belowdisplayskip{2pt}
\begin{align}\label{eq1}
\textbf H_k=\textbf G\bm \Phi^H \textbf h_k,
\end{align}
}where $\textbf G=\textbf G_t^H\bm\Lambda^H\textbf G_r$. The terms mentioned are defined as follows:
\begin{itemize}
  \item $\textbf G_t=[\textbf g_1,...,\textbf g_N]\in\mathbbmss{C}^{L\times N}$ is the transmit field response vector(FRV) at the BS, where $L$ is the number of channel paths between the BS and RIS and $\textbf g_n \in \mathbbmss{C}^{L\times 1}$ is the transmit FRV between the $n$-th MA and the RIS for $n=1,...,N $.
  \item $\textbf g_n=[{\rm {e}}^{j\frac{2\pi}{\lambda}\textbf t_n^T\bm\rho_{n,1}},...,{\rm e}^{j\frac{2\pi}{\lambda}\textbf t_n^T\bm\rho_{n,L}}]^T$, where $\bm\rho_{n,l}=[{\rm sin}\theta_{n,l}{\rm cos}\phi_{n,l},{\rm cos}\theta_{n,l}]^T, \theta_{k,l}\in[0, 2\pi]$ and $\phi_{k,l}\in[0, 2\pi]$ are the elevation and azimuth angles of the $l$-th path, respectively, for $l=1,...,L$. And $\lambda$ is the wavelength.
  \item $\bm\Lambda={\rm diag}\{[\nu_1,...,\nu_L]\} \in \mathbbmss{C}^{L\times L}$, where $\nu_l$ is the complex response of the $l$-th path.
  \item $\textbf G_r\in\mathbbmss{C}^{L\times M}$ is the receive FRV of the RIS.
  \item $\bm \Phi={\rm diag}(\psi_1,...,\psi_M)\in \mathbbmss{C}^{M\times M}$ is the phase-shift matrix of the RIS, where $\psi_m$ is the reflection coeffecient (RC) of $m$-th RIS element for $m=1,...,M$.
  \item $\textbf h_k=[h_{k_1},...,h_{k_M}]^T \in \mathbbmss{C}^{M\times 1}$ is the channel response between the RIS and the $k$-th UT for $k=1,...,K$.
\end{itemize}

Besides, we consider following three assumptions for the feasible set of RC in this paper:
\begin{itemize}
\item Ideal RC(IRC):  It only restricts that the RC is peak-power constrained:
\setlength\belowdisplayskip{2pt}
\begin{align}\label{eq2}
\mathcal F_1=\{\psi_m| \mid\psi_m\mid^2\leq 1\}.
\end{align}
\item Continuous Phase Shifter(CPS): It is assumed that the strength of the reflection signal from each reflection element is maximized, thus $|\psi_m|^2=1$, which can be represented as:\setlength\abovedisplayskip{2pt}
\setlength\belowdisplayskip{2pt}
\begin{align}\label{eq3}
\mathcal F_2=\{\psi_m|\psi_m={\rm e}^{j\xi_m}, \xi_m\in[0,2\pi)\}.
\end{align}

\item Discrete Phase Shifter(DPS): In practice, the phase of RC has finite levels. We assume that the phase of $\psi_m$ only takes $\kappa$ arithmetic values, which can be represented as:\setlength\abovedisplayskip{2pt}
\setlength\belowdisplayskip{2pt}
\begin{align}\label{eq35}
{\mathcal F_3=\{\psi_m|\psi_n={\rm e}^{j\xi_m}, \xi_m\in\{0,\frac{2\pi}{\kappa},...,\frac{2\pi(\kappa-1)}{\kappa}\}\}}.
\end{align}
\end{itemize}

Denoted by $\textbf x\in \mathbbmss{C}^{N\times 1}$ the input of the downlink transmission, then the received signal at $k$-th UT is expressed as:
{\setlength\abovedisplayskip{2pt}
\setlength\belowdisplayskip{2pt}
\begin{align}\label{eq4}
y_k=\textbf H_k^H\textbf x+n_k,
\end{align}
}where $n_k$ is the circularly symmetric complex Gaussian noise with zero mean and covariance $\sigma_k^2$. And $\textbf x=\sum_{k=1}^{K}\textbf w_k x_k$, where $x_k\in\mathbbmss{C}$ denotes the transmit data symbol to $k$-th UT with $\textbf w_k\in\mathbbmss{C}^{N\times 1}$ being corresponding transmit beamforming vector. It is assumed that all the $x_k$ are independent random variables with zero mean and unit variance. The received signal-to-interference-plus-noise ratio(SINR) at $k$-th UT is expressed as:
{\setlength\abovedisplayskip{2pt}
\setlength\belowdisplayskip{2pt}
\begin{align}\label{eq5}
\gamma_k=\frac{|\textbf h_k^H \bm\Phi \textbf G_r^H\bm\Lambda\textbf G_t\textbf w_k|^2}{\sum_{i\neq k}|\textbf h_k^H \bm\Phi \textbf G_r^H\bm\Lambda\textbf G_t\textbf w_i|^2+\sigma_k^2},
\end{align}
}Note that $\textbf G_t$ depends on the position of MAs. Then the sum-rate can be expressed as:
{\setlength\abovedisplayskip{2pt}
\setlength\belowdisplayskip{2pt}
\begin{align}\label{eq6}
R=\sum_{k=1}^{K}{\rm log}(1+\gamma_k).
\end{align}
}
\subsection{Problem Formulation}
This paper aims to improve the sum-rate of the system by jointly optimizing the transmit beamforming, the RC of RIS and the positions of MAs, subject to appropriate constraints. The transmit power constraint of BS is $\sum_{k=1}^{K}|\textbf w_k|^2\leq P_{max}.$

A minimum distance $D$ is required between each pair of MAs in order to avoid the coupling effect between the antennas in the transmit region. Thus we can formulate the sum-rate optimization problem as:
\begin{subequations}\label{P_1}
\begin{align}
{\mathcal{P}}_{\text{1}}:&~\max_{{\textbf{W}},{\textbf T},{\bf \Phi}}~{R}\label{P_X_Obj}\\
{\text{s.t.}}&~\sum\nolimits_{k=1}^{K}\textbf{w}_{k}^H\textbf{w}_{k}\leq P_{\max },\label{P_X_Cons1}\\
&~\psi_m\in\mathcal{F},\forall m=1,...,M, \label{P_X_Cons2}\\
&~\textbf t_n\in\mathcal{C}, \forall n=1,...,N, |\textbf t_n-\textbf t_{n'}|\leq D, n\neq n'\label{P_X_Cons3}
\end{align}
\end{subequations}where $\textbf W=[\textbf w_1,...,\textbf w_K]\in\mathbbmss{C}^{N\times K}$,$\mathcal{F}\in\{\mathcal F_1,\mathcal F_2,\mathcal F_3\}$ and $\textbf T=[\textbf t_1,...,\textbf t_N]\in\mathbbmss{R}^{2\times N}$. The above problem is difficult to tackle directly due to the fact that the objective function \eqref{P_X_Obj} is not convex and the constraints \eqref{P_X_Cons3} is not a convex set. Moreover, the coupling among $\textbf W$, $\bm\Phi$ and $\textbf T$ makes this problem more challenging to solve.

\section{PROPOSED ALGORITHM }
In this section, we present a joint optimization algorithm to solve problem \eqref{P_1}. Firstly, \eqref{P_X_Obj} is simplified into a more trackable form by invoking the FP method, i.e., the Lagrangian dual transform and quadratic transform proposed in \cite{Shen2018}. Then, the variables to be optimized are updated in an alternating manner, with other variables fixed. We present three solutions for different constraint conditions in \eqref{P_X_Cons2}.
\subsection{Optimizing beamforming matrix \textbf W}
We firstly introduce two auxiliary variable $\bm\alpha=[\alpha_1,...,\alpha_K]$ and $\bm\beta=[\beta_1,...,\beta_K]$ to turn \eqref{P_X_Obj} into a convex equivalent form. Then the problem can be written as:
\begin{subequations}\label{P_2}
\begin{align}
{\mathcal{P}}_{\text{2}}:&~\max_{{\textbf{W}},{\textbf T},{\bm \Phi},{\bm \alpha},{\bm \beta}}~{R'=\sum_{k=1}^{K}{\rm log}(1+\alpha_k)}-\sum_{k=1}^{K}\alpha_{k}\nonumber\\
&~+\sum_{k=1}^{K}(1+\alpha_k)[2\Re\{\beta_k^* A_k\}-|\beta_k|^2B_k]\\
{\text{s.t.}}&~\eqref{P_X_Cons1},~\eqref{P_X_Cons2},~\eqref{P_X_Cons3},\\
&~\alpha_k>0,\beta_k\in\mathbbmss{C},k\in\mathcal{K}.
\end{align}
\end{subequations}where $ A_k=\textbf h_k^H \bm\Phi \textbf G_r^H\bm\Lambda\textbf G_t\textbf w_k$ and $ B_k=\sigma_k^2+\sum_{i=1}^{K}|\textbf h_k^H \bm\Phi \textbf G_r^H\bm\Lambda\textbf G_t\textbf w_i|^2$. In \eqref{P_2}, when $\textbf W$, $\bm\Phi$ and $\textbf T$ hold fixed, the optimal $\alpha_k$ and $\beta_k$ are given by
{\setlength\abovedisplayskip{2pt}
\setlength\belowdisplayskip{2pt}
\begin{align}\label{eq10}
\alpha_k^\circ=\gamma_k
\end{align}
}
{\setlength\abovedisplayskip{2pt}
\setlength\belowdisplayskip{2pt}
\begin{align}\label{eq11}
\beta_k^\circ=\frac{\sqrt{1+\alpha_k} A_k}{B_k}
\end{align}
}

After updating $\alpha_k$ and $\beta_k$, \eqref{P_2} becomes a standard convex quadratic optimization problem whose solution is \cite{Shen2018} when $\bm\Phi$ and $\textbf T$ are fixed. The optimal $\textbf w_k$ is given by
{\setlength\abovedisplayskip{2pt}
\setlength\belowdisplayskip{2pt}
\begin{align}\label{eq12}
\textbf w_k^\circ=\sqrt{1+\alpha_k}\beta_k(\lambda_0 \textbf I+\sum_{i=1}^{K}|\beta_i|^2\textbf C_i)^{-1}\textbf G_t^H\bm\Lambda\textbf G_r\bm\Phi^H\textbf h_k
\end{align}
}where $\textbf C_i=\textbf G_t^H\bm\Lambda\textbf G_r\bm\Phi^H\textbf h_i\textbf h_i^H \bm\Phi \textbf G_r^H\bm\Lambda\textbf G_t$ and $\lambda_0$ is the dual variable introduced for the power constraint, which is optimally determined by
{\setlength\abovedisplayskip{2pt}
\setlength\belowdisplayskip{2pt}
\begin{align}\label{eq13}
\lambda_0^\circ=min\{\lambda_0\geq 0: \sum_{k=1}^{K}|\textbf w_k|^2\leq P_{max}\}.
\end{align}
}The dual variable $\lambda_0$ is chosen, such that the complementarity slackness, i.e., $\lambda_0\left(\sum_{k=1}^{K}|\textbf w_k|^2- P_{max}\right)$, is satisfied. It can be seen that $\sum_{k=1}^{K}|\textbf w_k|^2$ is a monotonic decreasing function with $\lambda_0\geq0$. Thus, we can find $\lambda_0$ via bisection-based search.
\subsection{Optimizing Reflection Response Matrix  \texorpdfstring{$\bm\Phi$ }{2}}
With $\textbf W, \alpha$ and $\beta$ fixed, the problem \eqref{P_2} can be expressed as:
\begin{subequations}\label{P_3}
\begin{align}
{\mathcal{P}}_{\text{3}}:&~\max_{{\textbf T},{\bf \Phi}}~{R''=\sum_{k=1}^{K}\left(1+\alpha_k\right) \frac{\gamma_k}{1+\gamma_k}}\\
{\text{s.t.}}&~\eqref{P_X_Cons2},~\eqref{P_X_Cons3}.
\end{align}
\end{subequations}
Simiarly, we introduce an auxiliary variable $\bm\epsilon=[\epsilon_1,...,\epsilon_K]$ and optimize $\bm\epsilon$ and $\bm\Phi$ alternatively. Then \eqref{P_3} can be translated to the following problem based on the quadratic transform proposed in \cite{Shen20182}:
\begin{subequations}\label{P_4}
\begin{align}
{\mathcal{P}}_{\text{4}}:&~\max_{{\textbf T},{\bf \Phi},{\bm\epsilon}}~{f\left(\textbf T,\bm\Phi,\bm\epsilon\right)}\label{P_O_3}\\
{\text{s.t.}}&~\eqref{P_X_Cons2},~\eqref{P_X_Cons3}.
\end{align}
\end{subequations}where the translated objective function is
{\setlength\abovedisplayskip{2pt}
\setlength\belowdisplayskip{2pt}
\begin{align}\label{eq16}
&~f\left(\textbf T,\bm\Phi,\bm\epsilon\right)=\sum_{k=1}^{K}[2\sqrt{1+\alpha_k}\Re\{\epsilon_k^H\bm\varphi^H\textbf{S}_k^H\textbf G_r^H\bm\Lambda\textbf G_t\textbf w_k\}\nonumber\\
&~-|\epsilon_k|^2\sigma_k^2+|\epsilon_k|^2\sum_{i=1}^{K}|\bm\varphi^H\textbf{S}_k^H \textbf G_r^H\bm\Lambda\textbf G_t\textbf w_i|^2]
\end{align}
}where $\bm\varphi=diag(\bm\Phi^H)\in \mathbbmss{C}^{M\times 1},\textbf{S}_k={\rm diag}(\textbf h_k)\in\mathbbmss{C}^{M\times M}$. When $\bm\Phi$ hold fixed, the optimal $\bm\epsilon$ can be calculated by setting $\partial f/\partial \epsilon_k$ to zero, which can be expressed as:
{\setlength\abovedisplayskip{2pt}
\setlength\belowdisplayskip{2pt}
\begin{align}\label{eq17}
\epsilon_k^\circ=\frac{\sqrt{1+\alpha_k}\bm\varphi^H\textbf{S}_k^H\textbf G_r^H\bm\Lambda\textbf G_t\textbf w_k}{\sum_{i=1}^{K}|\bm\varphi^H\textbf{S}_k^H \textbf G_r^H\bm\Lambda\textbf G_t\textbf w_i|^2+\sigma_k^2}
\end{align}
}

With $\bm\epsilon$ fixed, the objective function can be written as:
{\setlength\abovedisplayskip{2pt}
\setlength\belowdisplayskip{2pt}
\begin{align}\label{eq18}
f_2(\bm\varphi)=-\bm\varphi^H\textbf U\bm\varphi+2\Re\{\textbf V^H\bm\varphi\}
\end{align}
}where
{\setlength\abovedisplayskip{2pt}
\setlength\belowdisplayskip{2pt}
\begin{align}
\textbf U=\sum_{k=1}^{K}|\epsilon_k|^2(\sum_{i=1}^{K}\textbf{S}_k^H\textbf G_r^H\bm\Lambda\textbf G_t\textbf w_i\textbf w_i^H\textbf G_t^H \bm\Lambda^H\textbf G_r\textbf S_k),  \nonumber
\end{align}
}
{\setlength\abovedisplayskip{2pt}
\setlength\belowdisplayskip{2pt}
\begin{align}\label{eq19}
\textbf V=\sum_{k=1}^{K}\epsilon_k^H\textbf{S}_k^H\textbf G_r^H\bm\Lambda\textbf G_t\textbf w_k  \nonumber
\end{align}
}

The function \eqref{eq18} is a quadratic concave function of $\bm\varphi$ since $\textbf U$ is a positive-definite matrix. If $\mathcal{F}=\mathcal{F}_1$, then the optimization of $\bm\varphi$ is a convex problem due to the convexity of $\mathcal{F}_1$. Then the problem can be transformed to its dual problem via Lagrange dual decomposition:
\begin{subequations}\label{P_5}
\begin{align}
{\mathcal{P}}_{\text{5}}:&~\max_{{\bm\varphi}}~{f_3(\bm\varphi,\bm\eta)=f_2\left(\bm\varphi\right)-\sum_{m=1}^{M}\eta_m(\bm\varphi^H\textbf e_m\textbf e_m^H\bm\varphi-1)}\\
{\text{s.t.}}&~\psi_m\in\mathcal{F}_1,\forall m=1,...,M, \label{P_X_Cons21}
\end{align}
\end{subequations}where $\textbf e_m \in \mathbbmss{R}^{M\times1}$ is an elementary vector with a one at the $m$-th position, $\bm\eta=[\eta_1,...,\eta_M]$ is the dual variable for the constraint of the $m$-th RIS element, i.e., $\bm\varphi^H\textbf e_m\textbf e_m^H\bm\varphi\leq1$. And the optimal $\bm\varphi$ can be obtained by setting $\partial f_3/\varphi$ to zero, which can be expressed as:
{\setlength\abovedisplayskip{2pt}
\setlength\belowdisplayskip{2pt}
\begin{align}\label{eq195}
\bm\varphi=(\sum_{m=1}^{M}\eta_m\textbf e_m\textbf e_m^H+\textbf U)^{-1}\textbf V
\end{align}
}

Therefore, the optimal dual variable vector $\bm\eta$ can be determined according to the constraints in \eqref{P_X_Cons21} via ellipsoid method \cite{Guo2019}.

If $\mathcal{F}=\mathcal{F}_2$ or $\mathcal{F}=\mathcal{F}_3$, then the optimization of $\bm\varphi$ is not a convex problem due to the non-convexity of $\mathcal{F}_2$. It can be written as:
\begin{subequations}\label{P_6}
\begin{align}
{\mathcal{P}}_{\text{6}}:&~\max_{{\bm\varphi}}~{f_2\left(\bm\varphi\right)}\label{P_O_5}\\
{\text{s.t.}}&~\psi_m\in\mathcal{F}_2,\forall m=1,...,M, \label{P_X_Cons22}
\end{align}
\end{subequations}

Thus, we optimize $\bm\varphi$ via minimum-maximum (MM) method, which has the following iterative form:
{\setlength\abovedisplayskip{2pt}
\setlength\belowdisplayskip{2pt}
\begin{align}\label{eq20}
\textbf Q_\tau(\bm\varphi)=-\bm\varphi^H\bm\Omega\bm\varphi+2\Re\{\bm\varphi^H(\bm\Omega-\textbf U)\bm\varphi_{\tau}\}\nonumber\\
-\bm\varphi_\tau^H(\bm\Omega-\textbf U)\bm\varphi_\tau+2\Re\{\textbf V^H\bm\varphi\}
\end{align}
}where $\textbf Q_\tau$ is a base value of \eqref{P_O_5} for the in the $\tau$-th iteration, $\bm\Omega=\lambda_{max}I$ and $\lambda_{max}$ is the maximum eigenvalue of $\textbf U$. Since the contraint \eqref{P_X_Cons22} guarantees that $\bm\varphi^H\bm\varphi=M$, $\bm\varphi^H\bm\Omega\bm\varphi$ and $\bm\varphi_\tau^H(\bm\Omega-\textbf U)\bm\varphi_\tau$ are not related to the value of $\bm\varphi$. If $\mathcal{F}=\mathcal{F}_2$, it is obvious that the optimal $\bm\varphi$ in the $\tau$-th iteration can be expressed as:
{\setlength\abovedisplayskip{2pt}
\setlength\belowdisplayskip{2pt}
\begin{align}\label{eq21}
{\bm\varphi_{\tau+1}={\rm e}^{j\angle\{(\bm\Omega-\textbf U)\bm\varphi_\tau-\textbf V\}}}.
\end{align}
}

If $\mathcal{F}=\mathcal{F}_3$, the optimal phase of $m$-th element of $\bm\varphi$ in the $\tau$-th iteration can be expressed as:
{\setlength\abovedisplayskip{2pt}
\setlength\belowdisplayskip{2pt}
\begin{align}\label{eq216}
{{\xi_{m_{\tau+1}}}={\min_{\xi_m\in\{0,...,\frac{2\pi(\kappa-1)}{\kappa}\}}~{|\xi_m-\angle\{(\bm\Omega-\textbf U)\bm\varphi_\tau-\textbf V\}}|}}.
\end{align}
}
\subsection{Optimizing MA Position Matrix \textbf T}
With other variables fixed, we translate the problem \eqref{P_3} to the following problem by introducing an auxiliary $\bm\delta=[\delta_1,...,\delta_K]$.
\begin{subequations}\label{P_7}
\begin{align}
{\mathcal{P}}_{\text{7}}:&~\max_{{\textbf T},{\bm\delta}}~{f_4(\textbf T)=\sum_{k=1}^{K}2\sqrt{1+\alpha_k}\Re\{\delta_k^H\bm\Phi^H\textbf{S}_k^H\textbf{G}_r^H\bm\Lambda\textbf{G}_t\textbf{w}_k\}} \nonumber\\
&~{-|\delta_k|^2 (\sum_{i=1}^{K}|\bm\Phi^H\textbf{S}_k^H\textbf{G}_r^H\bm\Lambda\textbf{G}_t\textbf{w}_i|^2+\sigma_k^2) }\\
{\text{s.t.}}&~\eqref{P_X_Cons3}.
\end{align}
\end{subequations}Similar to \eqref{eq17}, the optimal $\bm\delta$ with fixed $\textbf T$ is:
{\setlength\abovedisplayskip{2pt}
\setlength\belowdisplayskip{2pt}
\begin{align}\label{eq215}
\delta_k^\circ=\frac{\sqrt{1+\alpha_k}\bm\varphi^H\textbf{S}_k^H\textbf G_r^H\bm\Lambda\textbf G_t\textbf w_k}{\sum_{i=1}^{K}|\bm\varphi^H\textbf{S}_k^H \textbf G_r^H\bm\Lambda\textbf G_t\textbf w_i|^2+\sigma_k^2}
\end{align}
}
Then the remaining problem is to optimize $\textbf T$. This optimization problem can be expressed as:
\begin{subequations}\label{P_8}
\begin{align}
{\mathcal{P}}_{\text{8}}:&~\max_{{\textbf T}}~f_5(\textbf T)=\sum_{k=1}^{K}2\sqrt{1+\alpha_k}\Re\{\textbf P^H_k\textbf G_t\textbf{w}_k\}\nonumber\\
&~ -\textbf P_k^H\textbf G_t\bm\Pi \textbf G_t^H\textbf P_k \label{P_O_6}\\
{\text{s.t.}}&~\eqref{P_X_Cons3}.
\end{align}
\end{subequations}where $\textbf P_k=\Lambda^H\textbf G_r{\textbf S_k}\bm\varphi\delta_k$ and $\bm\Pi=\sum_{i=1}^{K}\textbf w_i\textbf w_i^H$. The objective function \eqref{P_O_6} can also be written as

$f_4(\textbf T)=\sum_{k=1}^{K}2\sqrt{1+\alpha_k}\Re\left\{\sum_{l}^{L}{P^*_{k_l}}{\rm e}^{j\frac{2\pi}{\lambda}\textbf t_n^T\bm\rho_{n,l}}{w_{k_n}} \right\}
-\sum_{n=1}^{N}\sum_{l=1}^{L}\sum_{n'=1}^{N}\sum_{l'=1}^{L}P^*_{k_l}P_{k_{l'}}{\rm e}^{j\frac{2\pi}{\lambda}(\textbf t_n^T\bm\rho_{n,l}-\textbf t_{n'}^T\bm\rho_{n',l'})}\Pi_{n,n'}$
where $P_{k_l}$ is the $l$-th element of $\textbf P_k$, $w_{k_n}$ is the $n$-th element of $\textbf w_k$ and $\Pi_{n,m}$ is the $(n,m)$-th element of $\bm \Pi$. Due to the intractability of $f_4(\textbf T)$, we adopt a gradient descent (GD) algorithm with backtracking line search \cite{Zhang2017} to optimize $\textbf T$. The gradient values of $f_4(\textbf T)$ w.r.t. $\textbf t_n$ are calculated as follows \cite{Hjorungnes2007}:
{\setlength\abovedisplayskip{2pt}
\setlength\belowdisplayskip{2pt}
\begin{align}\label{eq22}
&~\nabla_{\textbf t_n}f_4=\sum_{k=1}^{K}\sqrt{1+\alpha_k}C_k
+E_k
+F_k
\end{align}
}where $C_k=\sum_{l=1}^{L}\frac{-4\pi\rho_{n,l}}{\lambda}|P^*_{k_l}w_{k_n}|{\rm sin}(\frac{2\pi}{\lambda}\textbf t_n^T\rho_{n,l}+\angle{P^*_{k_l}w_{k_n}})$, $E_k=\sum_{n'\neq n}^{N}\sum_{l=1}^{L}\sum_{l'=1}^{L}\frac{4\pi\rho_{n,l}}{\lambda}|P^*_{k_l}P_{k_{l'}}\Pi_{n',n}\hfill{ {\rm e}^{-j\frac{2\pi}{\lambda}\textbf t_{n'}^T\bm\rho_{n',l'}}|}$\\${\rm sin}(\frac{2\pi}{\lambda}\textbf t_n^T\bm\rho_{n,l}+\angle{P^*_{k_l}P_{k_{l'}}\Pi_{n',n}{\rm e}^{-j\frac{2\pi}{\lambda}\textbf t_{n'}^T\bm\rho_{n',l'}}})$ and ${F_k=}$\\${\sum_{l=1}^{L}\sum_{l'=1}^{L}\frac{2\pi(\rho_{n,l}-\rho_{n,l'})}{\lambda}|P^*_{k_l}P_{k_{l'}}\Pi_{n,n}|{\rm sin}(\frac{2\pi}{\lambda}\textbf t_n^T(\bm\rho_{n,l}-\bm\rho_{n,l'})}$\\$+\angle{P^*_{k_l}P_{k_{l'}}\Pi_{n,n}})$. The algorithm of optimizing $\textbf T$ is given in Algorithm 1, which is guaranteed to converge since the sum-rate is upper bounded. The proposed joint optimization algorithm is summarized in Algorithm 2, which is also guaranteed to converge to a stationary solution of Problem $\mathcal{P}_1$.
\vspace{-11pt}
\begin{algorithm}
    \caption{GD-Based Algorithm for Optimizing \textbf T}
    \label{Algorithm1}
    Initialize $\textbf{T}^0=[\textbf t_1^0,...,\textbf t_N^0]$, step size $\mu_{0}$, the minimum step size $\mu_{min}$, the iteration index $q=0$ and the maximum iteration number $q_{max}$.
    \Repeat{\textnormal{converge or reach the predefined maximum number of iterations $q_{max}$}}{
    \For{n=1:N}{
    Calculate  $\nabla_{\textbf t_n}f_4$ by \eqref{eq22} and set $\mu=\mu_0$;\\
    \Repeat{\textnormal{\eqref{P_X_Cons3}\& $f_4( \hat t_n)>f_4( t_n^{q}) or \mu<\mu_{min}$}}{
      Calculate $\hat{\textbf t}_n=\textbf{t}_n^q+\mu\nabla_{\textbf{t}_n}{f_4}$;\\
      Set  $\mu=\mu/2$;
    }

   Set $q=q+1$.\\
}
}
\end{algorithm}
\vspace{-20pt}
\begin{algorithm}
    \caption{FP-Based Algorithm for Solving $\mathcal{P}_1$}
    \label{Algorithm2}
    Initialize $\{\textbf W^0,\bm\Phi^0,\textbf{T}^0\}$, the iteration index $r=0$ and the maximum iteration number $r_{max}$.

    \Repeat{\textnormal{converge or reach the predefined maximum number of iterations $r_{max}$}}{
    Update $\bm\alpha^r$ and $\bm\beta^r$ by \eqref{eq10} and \eqref{eq11} \\
    Calculate $\textbf W^{r+1}$ by \eqref{eq12}\\
    Update $\bm\epsilon^r$  by \eqref{eq17}\\
    \If {$\mathcal{F}=\mathcal{F}_1$}{
    Update $\bm\varphi^r$  by \eqref{eq195}}
    \ElseIf{$\mathcal{F}=\mathcal{F}_2$ or $\mathcal{F}=\mathcal{F}_3$}{
    Set $\tau=0$ and $\tau_{max}$ for MM method\\
    \Repeat{\textnormal{converge or reach $\tau_{max}$}}{
    Update $\bm\varphi_\tau$ by \eqref{eq21} or \eqref{eq216}\\
    Set $\tau=\tau+1$}
    Update  $\bm\varphi^r= \bm\varphi_\tau$ and obtain $\bm\Phi^{r+1}$
    }
   Update $\bm\delta$  by \eqref{eq215}\\
   Solve $\mathcal{P}_8$ by Algorithm 1, and obtain $\textbf{T}^{r+1}$

   Set $r=r+1$.\\
}
\end{algorithm}
\vspace{20pt}
\section{Numerical Results}

In this section, numerical results are provided to confirm the effectiveness of the proposed algorithm. In the simulation, we set $N=4$, $K=4$, $M=16$, $D=\frac{\lambda}{2}$ and $\sigma_k^2=-100$dBm($\forall k$). The distance between the BS and RIS is 50 m. The UTs are uniformly and randomly distributed around a center, which is 100 m away from RIS, with radius 10 m. Moreover, we have incorporated the free-space path loss model for this system. It is widely known that the signal reflected by the RIS suffered from the “double-fading” effect. Thanks to the recent advances in meta-materials, the reflection gain of the RIS elements is usually high. We assume $\varsigma=10$dB denotes the reflection gain of RIS. Then the equivalent free-space path loss model of $k$-th UT is given by $-10{\rm{log}}_{10}v=92.5+20{\rm{log}}_{10}[f_0({\rm{GHZ}})]+20{\rm{log}}_{10}[d_0({\rm{km}})]-\varsigma$ and $-10{\rm{log}}_{10}u_k=92.5+20{\rm{log}}_{10}[f_0({\rm{GHZ}})]+20{\rm{log}}_{10}[d_k({\rm{km}})]-\varsigma$, where $f_0=2$GHz is the carrier frequency ,$d_0$ is the distance from the BS to the RIS and $d_k$ is the distance from the RIS to the $k$-th UT. As for the channel model, we assume that $L=4$, $\nu_l\sim\mathcal{CN}(0,\frac{v}{L})(\forall l)$ and  $h_{k_m}\sim\mathcal{CN}(0,u_k)(\forall k,m)$. The elevation and azimuth angles are randomly set within $[0 \pi]$. We compare the performance of MA and FPA-based scheme, where the BS is equipped with a uniform linear array composed of $N$ FPAs spaced by $\frac{\lambda}{2}$. We also compare the performance of RIS optimization and fixed RIS case, where the RC of the RIS is initialized with random phase and unit amplitude.

\begin{figure}[!t]
\centering
\setlength{\abovecaptionskip}{0pt}
\includegraphics[height=0.29\textwidth]{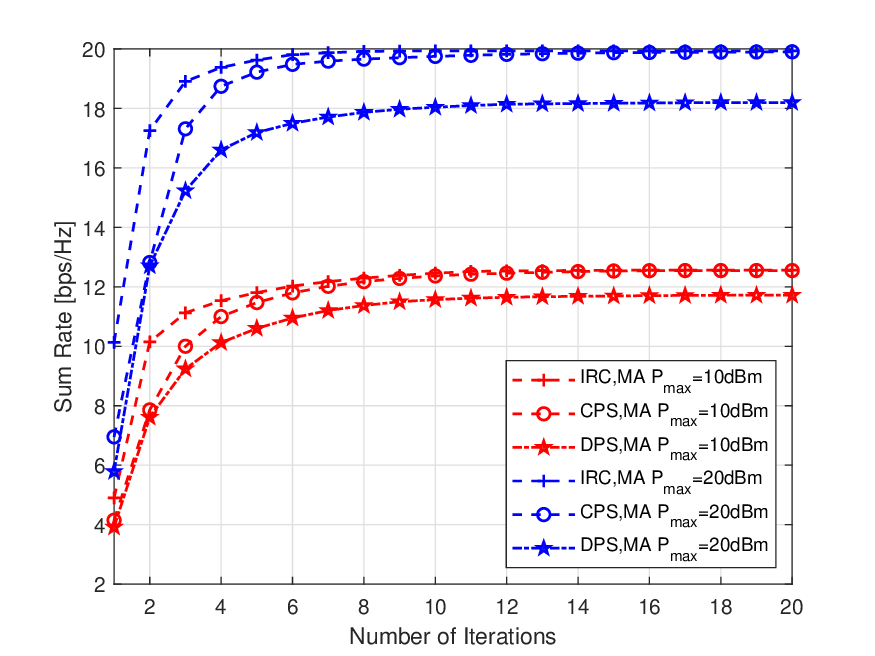}
\captionsetup{font={stretch=1.5}}
\caption{Sum-rate versus the number of iterations.}
\label{iter}
\vspace{-20pt}
\end{figure}

Firstly, we depict the convergence of the proposed methods for MAs in {\figurename} \ref{iter}. From this picture, it is observed that the sum-rate tends to converge to a stable value and all the three algorithms converge in about 12 iterations.

\begin{figure}[!t]
\centering
\setlength{\abovecaptionskip}{0pt}
\includegraphics[height=0.29\textwidth]{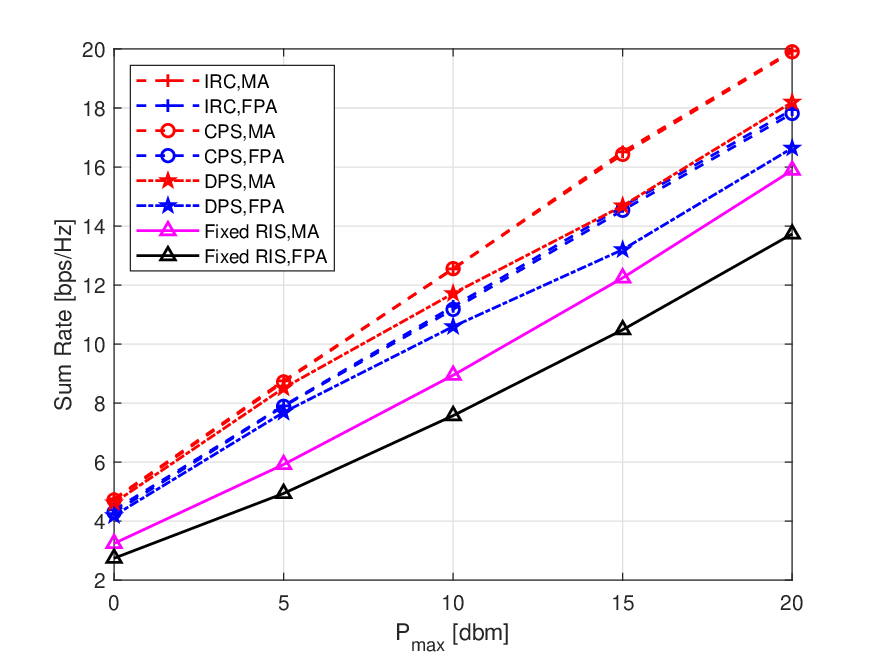}
\captionsetup{font={stretch=1.5}}
\caption{Sum-rate versus the power budget. $A=2\lambda$}
\label{pmax}
\vspace{-20pt}
\end{figure}

Then, we present the sum-rate versus the transmit power budget $P_{max}$ in {\figurename} \ref{pmax}. By comparing the performance of MA and FPA, it can be seen that the proposed algorithm achieves a larger sum-rate than the schemes with FPAs. Furthermore, we compare the performance of different RIS set. The IRC constraint have the best performance, but the sum-rate loss is negligible when it reduces to CPS constraint. It means that when the RIS is set with CPS constraint, the amplitude of the optimal RC is close to 1. The DPS constraint scheme achieves lower sum-rate than IRC and CPS since the inaccurate phase of RC.

\begin{figure}[!t]
\vspace{4pt}
\centering
\setlength{\abovecaptionskip}{0pt}
\includegraphics[height=0.29\textwidth]{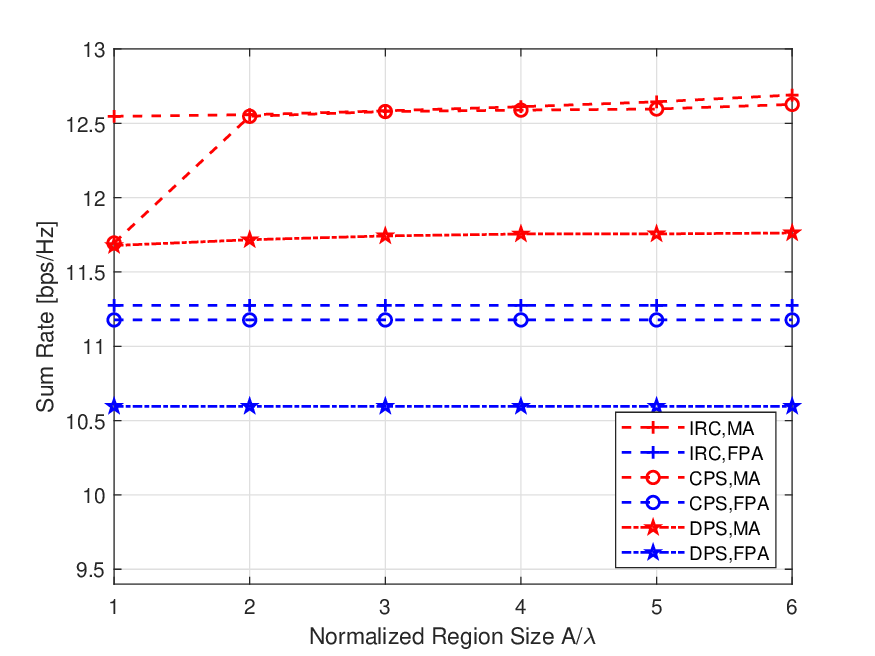}
\captionsetup{font={stretch=1.5}}
\caption{Sum-rate versus the normalized region size.}
\label{As}
\vspace{-20pt}
\end{figure}

Lastly, we show the sum-rate versus the normalized region size $A/\lambda$ when $P_{max}=10$dBm in {\figurename} \ref{As}. The achievable sum rate increases when the normalized region size gets larger. For CPS constraint, the performance tend to converge when $A/\lambda$ is over 3 while for IRC and DPS, the gain of region size is not obvious. This reveals that the optimal sum-rate performance for MA enbaled systems can be achieved with a finit transmit region and it is necessary to set a large enough region size for MA especially when the RIS is taken into account.

\section{Conclusion}
In this paper, we investigated the RIS-aided movable antenna enabled multiuser communication system with the aim of optimizing the sum-rate through beamforming, RC of RIS and the positions of MAs. We presented a FP-based algorithm to solve the joint optimization problem. Numerical results suggested that the RIS-aided MA enabled architecture outperforms conventional schemes in terms of sum-rate.

\end{document}